\documentclass[twocolumn,a4paper]{revtex4}
\usepackage{amsfonts,amsmath,amssymb,mathrsfs,dsfont,yfonts,bbm}
\begin{document}

\newcommand{\Supertwistor}{\Cset \mathrm{P}^{3|4}}
\newcommand{\Twistorspace}{\Cset \mathrm{P}^{3}}
\newcommand{\half}{\frac{1}{2}}
\newcommand{\diff}{\mathrm{d}}
\newcommand{\ra}{\rightarrow}
\newcommand{\Zset}{{\mathbb Z}}
\newcommand{\Cset}{{\,\,{{{^{_{\pmb{\mid}}}}\kern-.47em{\mathrm C}}}}}
\newcommand{\Rset}{{\mathrm{I}\!\mathrm{R}}}
\newcommand{\gra}{\alpha}
\newcommand{\grl}{\lambda}
\newcommand{\gre}{\epsilon}
\newcommand{\zb}{{\bar{z}}}
\newcommand{\mn}{{\mu\nu}}
\newcommand{\Acal}{{\mathcal A}}
\newcommand{\Rcal}{{\mathcal R}}
\newcommand{\Dcal}{{\mathcal D}}
\newcommand{\Mcal}{{\mathcal M}}
\newcommand{\Ncal}{{\mathcal N}}
\newcommand{\Kcal}{{\mathcal K}}
\newcommand{\Lcal}{{\mathcal L}}
\newcommand{\Scal}{{\mathcal S}}
\newcommand{\Wcal}{{\mathcal W}}
\newcommand{\Bcal}{\mathcal{B}}
\newcommand{\Ccal}{\mathcal{C}}
\newcommand{\Vcal}{\mathcal{V}}
\newcommand{\Ocal}{\mathcal{O}}
\newcommand{\Zcal}{\mathcal{Z}}
\newcommand{\Zb}{\overline{Z}}
\newcommand{\Urm}{{\mathrm U}}
\newcommand{\Srm}{{\mathrm S}}
\newcommand{\SO}{\mathrm{SO}}
\newcommand{\Sp}{\mathrm{Sp}}
\newcommand{\SU}{\mathrm{SU}}
\newcommand{\U}{\mathrm{U}}
\newcommand{\be}{\begin{equation}}
\newcommand{\ee}{\end{equation}}
\newcommand{\Comment}[1]{{}}
\newcommand{\tQ}{\tilde{Q}}
\newcommand{\tq}{{\tilde{q}}}
\newcommand{\trho}{\tilde{\rho}}
\newcommand{\tphi}{\tilde{\phi}}
\newcommand{\Qcal}{\mathcal{Q}}
\newcommand{\tmu}{\tilde{\mu}}
\newcommand{\dbar}{\bar{\partial}}
\newcommand{\p}{\partial}
\newcommand{\eg}{{\it e.g.\;}}
\newcommand{\ie}{{\it i.e.\;}}
\newcommand{\Tr}{\mathrm{Tr}}
\newcommand{\twistor}{\Cset \mathrm{P}^{3}}
\newcommand{\note}[2]{{\footnotesize [{\sc #1}}---{\footnotesize   #2]}}
\newcommand{\CL}{\mathcal{L}}
\newcommand{\CJ}{\mathcal{J}}
\newcommand{\CA}{\mathcal{A}}
\newcommand{\CH}{\mathcal{H}}
\newcommand{\CD}{\mathcal{D}}
\newcommand{\CE}{\mathcal{E}}
\newcommand{\CQ}{\mathcal{Q}}
\newcommand{\CB}{\mathcal{B}}
\newcommand{\CC}{\mathcal{C}}
\newcommand{\CO}{\mathcal{O}}
\newcommand{\CT}{\mathcal{T}}
\newcommand{\CI}{\mathcal{I}}
\newcommand{\CN}{\mathcal{N}}
\newcommand{\CS}{\mathcal{S}}
\newcommand{\CM}{\mathcal{M}}

\parskip 11pt

\title{\Large {\bf Nonunitary Lagrangians and unitary non-Lagrangian conformal~field~theories}} 
\author {Matthew Buican and Zoltan Laczko} 
\affiliation{CRST and School of Physics and Astronomy \\ Queen Mary University of London, London E1 4NS, UK\\ }

\begin{abstract}
In various dimensions, we can sometimes compute observables of interacting conformal field theories (CFTs) that are connected to free theories via the renormalization group (RG) flow by computing protected quantities in the free theories. On the other hand, in two dimensions, it is often possible to algebraically construct observables of interacting CFTs using free fields without the need to explicitly construct an underlying RG flow. In this note, we begin to extend this idea to higher dimensions by showing that one can compute certain observables of an infinite set of unitary strongly interacting four-dimensional $\CN=2$ superconformal field theories (SCFTs) by performing simple calculations involving sets of non-unitary free four-dimensional hypermultiplets. These free fields are distant cousins of the Majorana fermion underlying the two-dimensional Ising model and are not obviously connected to our interacting theories via an RG flow. Rather surprisingly, this construction gives us Lagrangians for particular observables in certain subsectors of many \lq\lq non-Lagrangian" SCFTs by sacrificing unitarity while preserving the full $\CN=2$ superconformal algebra. As a byproduct, we find relations between characters in unitary and non-unitary affine Kac-Moody algebras. We conclude by commenting on possible generalizations of our construction.
\end{abstract}
\maketitle

\section*{Introduction}
Free fields in two spacetime dimensions are versatile: operators, correlation functions, and partition functions of interacting conformal field theories (CFTs) can often be constructed algebraically from free bosons via the Coulomb gas formalism, and the simplest unitary minimal model---the Ising model---has a free Majorana fermion underlying it (see \cite{DiFrancesco:1997nk} for a review). Free fields in higher dimensions seem less powerful: in order to have something useful to say about an interacting CFT, one must usually labor to connect such free fields to the CFT in question through a suitably \lq\lq smooth" path in the space of couplings \footnote{We will not make this notion of smoothness precise here, except to say that, at the very least,  there should not be any accidental symmetries along the resulting renormalization group (RG) flow that obscure the observable one would like to compute.}.

However, one may hope to overcome these obstacles in $d>2$ spacetime dimensions whenever there are relations between quantum field theories (QFTs) in $d$ dimensions and QFTs in 2D. In the case of 4D superconformal field theories (SCFTs) with at least $\mathcal{N}=2$ supersymmetry (SUSY), one such relation was given in \cite{Beem:2013sza}: the sector of so-called \lq\lq Schur" operators of the 4D SCFT (briefly reviewed in the supplementary material) is isomorphic to a 2D chiral algebra living on a plane, $\mathcal{P}\subset\mathbb{R}^4$. On the chiral algebra side of this relation, one of the most basic quantities we can compute is the torus partition function
\begin{equation}\label{toruspfn}
Z(x,q)\equiv q^{-{c_{2d}\over24}}\Tr\ x^{M^{\perp}}q^{L_0}~,
\end{equation}
where the trace is over the Hilbert space of states associated with the chiral algebra, $c_{2d}$ is the chiral algebra central charge, $M^{\perp}=j_1-j_2$ is the spin transverse to $\mathcal{P}$ ($j_{1,2}$ are Cartans of $SO(4)$), $q\in\mathbb{C}$ is a fugacity, and $L_0$ gives the holomorphic scaling dimension, $h$. On the 4D side of the relation, \eqref{toruspfn} is mapped to a particular refined Witten index, called the Schur index \cite{Gadde:2011uv} (see the supplementary material for further details), that counts the Schur operators weighted by certain quantum numbers
\begin{equation}\label{SchurI}
\mathcal{I}_{\rm S}(q)\equiv q^{c_{4d}\over2}{\rm Tr}_{\mathcal{H}}(-1)^Fq^{E-R}=Z(-1,q)~,
\end{equation}
where $c_{4d}$ is the 4D $c$ central charge, $F$ is fermion number, $E$ is the scaling dimension, and $R$ is the $su(2)_R$ weight (clearly, the holomorphic scaling dimension satisfies $h=E-R$ while $c_{2d}=-12c_{4d}$ \cite{Beem:2013sza}). Note that both \eqref{toruspfn} and \eqref{SchurI} can be refined by additional flavor fugacities (i.e., fugacities for symmetries that commute with $\mathcal{N}=2$ SUSY in 4D), but such modifications will not play a role in our discussion below.

While we believe that many of the ideas we will present are quite broadly applicable (with suitable modifications), in this note we specialize to a particular infinite set of strongly coupled SCFTs whose simplest member is the so-called $(A_1, D_4)$ theory \footnote{This theory was originally discovered in \cite{Argyres:1995xn}, but we use the naming conventions of \cite{Cecotti:2010fi}.}. In this class, the manipulations we use are particularly simple.

The Schur index for the $(A_1, D_4)$ theory was computed in \cite{Buican:2015ina,Buican:2015hsa,Cordova:2015nma,Buican:2015tda} and was shown to equal the vacuum character of $\widehat{su(3)}_{-{3\over2}}$ (as conjectured in \cite{RastelliSeminar}). More recently, the authors of \cite{Xie:2016evu} proposed that this unflavored Schur index takes the following simple form
\begin{equation}\label{XYY}
I_S^{(A_1, D_4)}(q)=q^{1\over3}{\rm P.E.}\left(8{q\over1-q^2}\right)\equiv q^{1\over3}{\rm Exp}\left(8\sum_{n=1}^{\infty}{1\over n}{q^n\over1-q^{2n}}\right)~,
\end{equation}
and this formula was proven in \cite{Kac:2017} (see also the discussion in \cite{Creutzig:2017qyf}) to be equivalent to the vacuum character of $\widehat{su(3)}_{-{3\over2}}$ \footnote{The authors of \cite{Song:2017oew} also argued in favor of \eqref{XYY} (and a generalization we will encounter below) using results in \cite{Buican:2015ina,Song:2015wta}.}. Interestingly, under the rescaling $q\to q^{1\over2}$, \eqref{XYY} reduces to
\begin{equation}\label{A1D4freehyper}
I_S^{(A_1, D_4)}(q^{1\over2})=q^{1\over6}{\rm P.E.}\left(8{q^{1\over2}\over1-q}\right)=\left(I_S^{\rm half-hyper}\right)^8~,
\end{equation}
where the righthand side (RHS) is just the index of eight free half-hypermultiplets (i.e., the $T_2$ theory \cite{Gaiotto:2009we}) or, equivalently in 2D, the vacuum character of four symplectic bosons.

While the derivation in \cite{Kac:2017} proved \eqref{XYY} \footnote{More precisely, the authors of \cite{Kac:2017} proved that the vacuum character of $\widehat{su(3)}_{-{3\over2}}$ is given by the RHS of \eqref{XYY}.} along with various generalizations we will encounter below, we would like to give a physical argument for why this index is so closely related to the index of free fields. One hint comes from the study in \cite{Buican:2017fiq} (building on \cite{Buican:2014hfa}) that shows the $(A_1, D_4)$ theory plays a role in a particular $S$-duality that is reminiscent of the role played by free hypermultiplets in the $S$-duality of \cite{Argyres:2007cn}. Moreover, by thinking of \eqref{A1D4freehyper} as a manifestation of a weak-strong \lq\lq duality" \footnote{The relation in \eqref{A1D4freehyper} is not a duality in the truest sense of the word since it is a relation between the Schur sectors of two different theories.} we, in collaboration with T.~Nishinaka, speculated that this connection might be related to modularity \cite{Buican:2017fiq}. 

As we will see below, this intuition is morally correct, although the free fields that are more closely related to modularity are actually non-unitary (wrong statistics) rather than the unitary fields appearing on the RHS of \eqref{A1D4freehyper}. A strong indication that this idea is correct comes from noting that \eqref{XYY} satisfies the modular differential equation \cite{Beem:2017ooy}
\begin{equation}\label{LMDED4}
\left(D_q^{(2)}-40\mathbb{E}_4\right)I_S^{(A_1, D_4)}=\left(D_q^{(2)}-40\mathbb{E}_4\right)\chi_0^{\widehat{su(3)}_{-{3\over2}}}=0~,
\end{equation}
where $D_q^{(2)}$ is a modular differential operator, and $\mathbb{E}_4$ is an Eisenstein series (we refer the interested reader to \cite{Beem:2017ooy} for more details). The characters of $\widehat{so(8)}_1$ satisfy the same modular differential equation \cite{Mathur:1988na}. Since $\widehat{so(8)}_1$ is unitary and has a representation in terms of eight free Majorana fermions, it is reasonable to imagine that the 4D ancestor of this theory is a non-unitary free theory (recall that, as discussed above, $c_{4d}=-{1\over12}c_{2d}$, so $c_{4d}<0$ in this case). Clearly, these free fields then reproduce some of the observables in the Schur sector of the $(A_1, D_4)$ SCFT.

\section{Modular $S$-transformations and an AKM relation}
In order to understand the modular properties of \eqref{XYY}, it is useful to re-write it as follows
\begin{equation}\label{XYY2}
I_S^{(A_1, D_4)}=2^{-4}{\theta_2(\tau)^{4}\over\eta(\tau)^4}~,
\end{equation}
where $q=e^{2\pi i\tau}$, $\eta(\tau)$ is the Dedekind eta function, and $\theta_i(\tau)$ are the Jacobi theta functions (see the supplementary material). Under a modular $S$-transformation, we have
\begin{equation}\label{modStrans}
\theta_2\left(-{1\over\tau}\right)=\sqrt{-i\tau}\theta_4(\tau)~, \ \ \ \eta\left(-{1\over\tau}\right)=\sqrt{-i\tau}\eta(\tau)~.
\end{equation}
In particular, we see that applying a modular $S$-transformation to \eqref{XYY2} yields
\begin{equation}\label{StransD4}
\CS\left(I_S^{(A_1, D_4)}\right)=2^{-4}{\theta_4(\tau)^{4}\over\eta(\tau)^4}=2^{-4}q^{-{1\over6}}\ {\rm P.E.}\left(-{8q^{1\over2}\over1-q}\right)~.
\end{equation}
We immediately recognize the expression on the RHS as also counting (with a $(-1)^F$ weighting) the $\widehat{so(8)}_1$ fields generated by acting on the $\widehat{so(8)}_1$ vacuum with the $h={1\over2}$ Majorana fermions in the ${\bf 8_{v}}$ representation, $\psi^I$ (where $I=1,\cdots, 8$) \cite{DiFrancesco:1997nk} \footnote{See the supplementary material for a brief description of the characters of $\widehat{so(2n)}_1$.}  (hence, this theory is related to eight decoupled Ising models). These fields have the following singular OPE
\begin{equation}\label{MajoranaOPE}
\psi^I(z)\psi^J(w)\sim{\delta^{IJ}\over z-w}~.
\end{equation}
At the level of characters, we have the relation
\begin{eqnarray}\label{modStransD4}
\CS\Big(\chi_0^{\widehat{su(3)}_{-{3\over2}}}\Big)&=&-{1\over2}\Big(\chi_0^{\widehat{su(3)}_{-{3\over2}}}+\chi_{-{1\over2},1}^{\widehat{su(3)}_{-{3\over2}}}+\chi_{-{1\over2},2}^{\widehat{su(3)}_{-{3\over2}}}\cr&-&\chi_{-{1\over2},3}^{\widehat{su(3)}_{-{3\over2}}}\Big)=2^{-4}\Big(\chi_0^{\widehat{so(8)}_1}-\chi_{{1\over2},v}^{\widehat{so(8)}_1}\Big)~,\ \ \ \ \ \
\end{eqnarray}
where, in the second equality, we have used our observation above and, in the first equality, we have used the modular $S$ matrix acting on the characters of the four admissible representations of $\widehat{su(3)}_{-{3\over2}}$
\begin{equation}\label{eq:S-matrix}
S_{\widehat{su(3)}_{-{3\over2}}}=-\frac{1}{2}
  \begin{pmatrix}
    1 & 1 & 1 & -1 \\
    1 & 1 & -1 & 1 \\
    1 & -1 & 1 & 1 \\
    -1 & 1 & 1 & 1
  \end{pmatrix}~.
\end{equation}
There are four admissible representations of $\widehat{su(3)}_{-{3\over2}}$ (the vacuum and three $h=-{1\over2}$ representations) and four representations of $\widehat{so(8)}_1$ (the vacuum and three $h={1\over2}$ representations), but in the latter case all four corresponding unrefined characters are finite, while in the former case only two linear combinations of unrefined characters are finite (the vacuum and the linear combination of $h=-{1\over2}$ characters, $\chi'^{\widehat{su(3)_{-{3\over2}}}}_{-{1\over2}}$, appearing in \eqref{modStransD4}). However, all the unrefined $h={1\over2}$ characters of $\widehat{so(8)}_1$ are equal (we denote the corresponding character $\chi'^{\widehat{so(8)}_1}_{1\over2}$), and we find the bijection of finite unrefined characters \footnote{Ultimately, the fact that there are only two characters transforming amongst each other under modular transformations in the case of our two AKM algebras is a consequence of the Jacobi quartic identity $\theta_3(\tau)^4=\theta_2(\tau)^4+\theta_4(\tau)^4$.}
\begin{equation}
\chi_0^{\widehat{su(3)}_{-{3\over2}}}\sim\chi'^{\widehat{so(8)}_1}_{1\over2}~, \ \ \ \chi'^{\widehat{su(3)_{-{3\over2}}}}_{-{1\over2}}\sim\chi_0^{\widehat{so(8)}_1}~,
\end{equation}
where the relations hold up to overall constants (see \cite{Mukhi:1989bp} for character relations between other pairs of unitary and non-unitary theories).
\section{A 4D interpretation}
We would like to give a 4D interpretation for the unitary $\widehat{so(8)}_1$ theory described in the previous section by using the relation discovered in \cite{Beem:2013sza} (although, apriori, it is not clear such an interpretation must exist). As discussed above, this theory should be non-unitary since
\begin{equation}
c_{\widehat{so(8)}_1}=4 \ \Rightarrow \ c_{4d}=-{1\over3}~.
\end{equation}
Moreover, from the results in \eqref{SchurI} and \eqref{StransD4}, we see that an obvious candidate for our 4D theory is a collection of 8 half-hypermultiplets with wrong statistics (i.e., a \lq\lq ghost" $T_2$ theory) \footnote{\label{evenOddD4} To get just the $\widehat{so(8)}_1$ vacuum module, we should restrict to composite Schur operators built from an even number of hypermultiplet scalars.}. Indeed, the $a$ and $c$ anomalies for such a theory are just minus the corresponding anomalies for the $T_2$ theory since the wrong statistics leads to an insertion of a factor of $-1$ in any quantum loop. In particular, we have
\begin{equation}
c_{4d}=-8\times c_{\rm half-hyper}=-{1\over3}~, \ \ \ a_{4d}=-8\times a_{\rm half-hyper}=-{1\over6}~.
\end{equation}
Note that $a_{4d}-c_{4d}$ is then consistent with the $q\to1$ \lq\lq Cardy" limit of the index \cite{DiPietro:2014bca,Buican:2015ina,Ardehali:2015bla,Beem:2017ooy}, and the full (unrefined) Schur index is precisely what we want (see the previous footnote).

To get a map of operators, the correspondence in \cite{Beem:2013sza} requires us to take 4D Schur operators, fix them in a plane (with coordinates $z,\bar z$), and then twist the global $\bar z$ conformal transformations with $su(2)_R$. Working in the cohomology of a particular supercharge, $\mathbbmtt{Q}$, then gives a map to 2D chiral algebra operators. This procedure is naturally implemented in the operator product expansion (OPE).

For the case at hand, we can build all Schur operators as arbitrary (non-vanishing) products of the $su(2)_R$ highest weight anti-commuting scalars of the non-unitary free hypermultiplets, $q^I$, and their derivatives. These fields are organized as $q^i=Q^i$ and $q^{i+4}=\tilde Q^i$ (with $i=1,\cdots,4$) and live in the following $su(2)_R$ doublets
\begin{equation}
\begin{pmatrix}
    Q^i \\
    \tilde Q^{i\dagger} 
\end{pmatrix}~, \ \ \ 
\begin{pmatrix}
    \tilde Q^i \\
    -Q^{i\dagger} 
\end{pmatrix}~.
\end{equation}
We can write a simple Lagrangian for this non-unitary theory (note that the spinors in the hypermultiplets commute while the scalars anti-commute)
\begin{equation}\label{LagrangianD4}
\mathcal{L}=-\int d^4\theta\left(q^{I\dagger}\Omega_{IJ}q^J\right)=\int d^4\theta\left(\tilde Q^{i\dagger}\delta_{ij}Q^j-Q^{i\dagger}\delta_{ij}\tilde Q^j\right)~,
\end{equation}
where we have defined
\begin{equation}
\Omega \equiv \begin{pmatrix}
    0_{4\times4} & 1_{4\times4}\\
    -1_{4\times 4} & 0_{4\times4} 
\end{pmatrix}~.
\end{equation}
Related Lagrangians have been considered in different contexts in \cite{Anninos:2011ui,Hertog:2017ymy}.

The non-vanishing singular OPEs are then (in an appropriate normalization to eliminate a common overall constant factor)
\begin{equation}
\tilde Q^{i\dagger}(x)Q^j(0)\sim{\delta^{ij}\over x^2}~, \ \ \ Q^{i\dagger}(x)\tilde Q^j(0)\sim-{\delta^{ij}\over x^2}~.
\end{equation}
According to the discussion in \cite{Beem:2013sza}, we should twist the hypermultiplets with vectors $u_i=(1,\bar z)$ having $su(2)_R$ indices $i=1,2$. In particular, we have twisted fields
\begin{eqnarray}
Q'^i(z,\bar z)&=&Q^i(z,\bar z)+\bar z\tilde Q^{i\dagger}(z,\bar z)~, \cr \tilde Q'^i(z,\bar z)&=&\tilde Q^{i}(z,\bar z)-\bar z Q^{i\dagger}(z,\bar z)~,
\end{eqnarray}
with the following singular OPEs
\begin{equation}\label{OPEs}
Q'^i(z,\bar z)Q'^j(0,0)\sim{\delta^{ij}\over z}~, \ \ \ \tilde Q'^i(z,\bar z)\tilde Q'^j(0,0)\sim{\delta^{ij}\over z}~.
\end{equation}
Passing to $\mathbbmtt{Q}$ cohomology gives the same OPEs as above (the identity operator is $\mathbbmtt{Q}$-closed but clearly cannot be $\mathbbmtt{Q}$-exact). In particular, we reproduce the free Majorana OPEs of \eqref{MajoranaOPE}.

The theory also has conserved currents sitting as level-two descendants in multiplets with Schur operators of the form
\begin{equation}\label{currentsD4}
\mu^{ij}=iQ^iQ^j~, \ \ \ \tilde\mu^{ij}=i\tilde Q^i\tilde Q^j~, \ \ \ \mu'^{ij}=iQ^i\tilde Q^j~,
\end{equation}
where $i,j=1,\cdots,4$. More covariantly, we can define these operators to form part of a 28-dimensional adjoint representation with $\mu^{IJ}=iq^Iq^J$ and $I,J=1,\cdots,8$ (this operator is anti-symmetric in $I$ and $J$). The charges arising from real currents sitting as descendants of linear combinations of the above satisfy an $so^*(8)\simeq so(6,2)$ Lie algebra, which is a real form of $so(8,\mathbb{C})$. On the other hand, the operators in \eqref{currentsD4} are related to currents that are not real. However, these currents give rise to charges that act in accordance with the reality condition in two dimensions
\begin{eqnarray}
\mu^{ij}&:&\delta Q^i\sim -Q^j~, \delta Q^j\sim Q^i~, \delta\tilde Q^{i\dagger}\sim-\tilde Q^{j\dagger}~, \delta\tilde Q^{j\dagger}\sim\tilde Q^{i\dagger}~,\nonumber\cr
\tilde\mu^{ij}&:&\delta\tilde Q^i\sim\tilde Q^j~, \delta\tilde Q^j\sim-\tilde Q^i~, \delta Q^{i\dagger}\sim Q^{j\dagger}~, \delta Q^{j\dagger}\sim -Q^{i\dagger}~,\nonumber\cr \mu'^{ij}&:&\delta\tilde Q^j\sim Q^i~, \delta Q^i\sim-\tilde Q^j~, \delta\tilde Q^{i\dagger}\sim Q^{j\dagger}~, \delta Q^{j\dagger}\sim -\tilde Q^{i\dagger}~.
\end{eqnarray}
Relabeling the moment maps with an adjoint index of $so(8)$, we obtain the following twisted OPE
\begin{equation}
\mu^A(z,\bar z)\mu^B(0)\sim{\delta^{AB}\over z^2}+if^{AB}_{\ \ \ C}{\mu^C(0,0)\over z}+\left\{\mathbbmtt{Q},\cdots\right\}~,
\end{equation}
where $f^{AB}_{\ \ \ C}$ are the structure constants of $so(8)$. Dropping the $\mathbbmtt{Q}$-exact terms then leads to the standard $\widehat{so(8)}_1$ current-current OPE. As a result, we see that a generalization of the procedure of \cite{Beem:2013sza} applied to our non-unitary 4D theory yields the desired unitary theory in 2D.

\section{Infinitely many generalizations}
One can imagine generalizing our discussion above in many directions. Here we choose the simplest direction: the $(A_1, D_4)$ theory is part of an infinite family of SCFTs called the $D_2[SU(2N+1)]$ theories \cite{Cecotti:2012jx,Cecotti:2013lda} (where $D_2[SU(3)]\equiv(A_1, D_4)$). The corresponding chiral algebras were found in \cite{Xie:2016evu} and were argued to be $\widehat{su(2N+1)}_{-{2N+1\over2}}$. The generalization of \eqref{XYY} is
\begin{equation}\label{XYYgen}
I_S^{D_2[SU(2N+1)]}(q)=q^{N(N+1)\over6}{\rm P.E.}\left(4N(N+1){q\over1-q^2}\right)~,
\end{equation}
and one finds that, upon taking $q\to q^{1\over2}$, the index \eqref{XYYgen} reduces to the index of $4N(N+1)$ free half-hypermultiplets.

For $N>1$, the modular properties of the theory are somewhat different. For example, the modular differential equation in \eqref{LMDED4} for the $N=1$ case becomes third order for all $N>1$. However, we can proceed as before and write
\begin{equation}
I_S^{D_2[SU(2N+1)]}=2^{-2N(N+1)}{\theta_2(\tau)^{2N(N+1)}\over\eta(\tau)^{2N(N+1)}}~.
\end{equation}
Then, performing a modular $S$-transformation yields
\begin{eqnarray}\label{StransGen}
\CS\left(I_S^{D_2[SU(2N+1)]}\right)=2^{-2N(N+1)}{\theta_4(\tau)^{2N(N+1)}\over\eta(\tau)^{2N(N+1)}}\cr=2^{-2N(N+1)}q^{-{N(N+1)\over12}}\ {\rm P.E.}\left(-{4N(N+1)q^{1\over2}\over1-q}\right)~.
\end{eqnarray}
This result generalizes the $N=1$ result discussed above, since we recognize \eqref{StransGen} as also counting (with a $(-1)^F$ weighting) the $\widehat{so(4N(N+1))}_1$ fields generated by acting on the $\widehat{so(4N(N+1))}_1$ vacuum with the $h={1\over2}$ Majorana fermion in the vector representation, $\psi^I$ (its singular self-OPE is the obvious generalization of \eqref{MajoranaOPE} with $I=1,\cdots, 4N(N+1)$).

The $\widehat{so(4N(N+1))}_1$ algebra has four representations for all $N$: the vacuum, the $h=1/2$ representation discussed above, and two $h=N(N+1)/4$ representations. The latter two representations have identitcal unrefined characters which we denote as $\chi'^{\widehat{so(4N(N+1))}_1}_{N(N+1)\over4}$ (for $N=1$, the last three unrefined characters are identical). On the other hand, the $\widehat{su(2N+1)}_{-{2N+1\over2}}$ algebra has three finite (linear combinations of) unrefined characters that transform into each other under modular transformations: one starting with $h=0$ (the vacuum), one starting with $h={2-N(N+1)\over4}$, and one starting with $h=-{N(N+1)\over4}$. It is straightforward to check that
\begin{eqnarray}\label{dictionaryGen}
\chi_0^{\widehat{su(2N+1)}_{-{2N+1\over2}}}&\sim&\chi'^{\widehat{so(4N(N+1))}_1}_{N(N+1)\over4}~,\cr\chi'^{\widehat{su(2N+1)}_{-{2N+1\over2}}}_{2-N(N+1)\over4}&\sim&\chi^{\widehat{so(4N(N+1))}_1}_{1\over2}~,\cr\cr\chi'^{\widehat{su(2N+1)}_{-{2N+1\over2}}}_{-{N(N+1)\over4}}&\sim&\chi^{\widehat{so(4N(N+1))}_1}_{0}~.
\end{eqnarray}
These results are simple consequences of the fact that our two chiral algebras satisfy the same modular differential equation for all $N$.

The 4D generalization of the $N=1$ case is straightforward. For example, we have that
\begin{equation}
c_{\widehat{so(4N(N+1))}_1}=2N(N+1)\Rightarrow c_{4d}=-{N(N+1)\over6}~.
\end{equation}
This anomaly is precisely what we expect for $4N(N+1)$ half-hypers with wrong statistics (i.e., $N(N+1)/2$ \lq\lq ghost" $T_2$ theories). Similarly, $a_{4d}$ and the superconformal index are compatible with this interpretation. In particular, our 4D Lagrangian is just the obvious generalization of \eqref{LagrangianD4}
\begin{equation}\label{LagGen}
\mathcal{L}=-\int d^4\theta\left(q^{I\dagger}\Omega_{IJ}q^J\right)=\int d^4\theta\left(\tilde Q^{i\dagger}\delta_{ij}Q^j-Q^{i\dagger}\delta_{ij}\tilde Q^j\right)~,
\end{equation}
where now $I=1,\cdots,4N(N+1)$. Note that the real flavor currents in 4D generate an $so^*(4N(N+1))$ algebra, but the $N(2N-1)$  Schur operators that are the generalizations of \eqref{currentsD4} give rise to the $\widehat{so(4N(N+1))}_1$ AKM algebra in 2D \footnote{To get just the $\widehat{so(4N(N+1))}_1$ vacuum module, we should restrict to composite Schur operators built from an even number of $q^I$.}.

\section{Discussion and Conclusions}
We have seen that the simple non-unitary 4D Lagrangian \eqref{LagGen} allows us (through manipulations in two dimensions) to exactly compute the unrefined Schur indices for the $D_2[SU(2N+1)]$ SCFTs. Clearly, we are also able to compute other (linear combinations of) characters of the associated chiral algebras via the, to our knowledge, novel mathematical identities in \eqref{dictionaryGen}. Based on known relations between chiral algebras in 2D and 3D QFT, it is reasonable to expect that aspects of the physics of the non-vacuum modules of the chiral algebras are captured by (worldvolumes of) 4D objects that have non-trivial braiding statistics as in \cite{Wang}. Indeed, there is considerable evidence that this intuition holds \cite{Cordova:2017mhb,Pan:2017zie}, and we hope to return to a detailed discussion of surface and line defects in our setup soon. In particular, the Lagrangian in \eqref{LagGen} seems to compute the Schur indices of the $D_2[SU(2N+1)]$ SCFTs in the presence of certain surface defects \footnote{Or, more precisely, it computes linear combinations of the usual Schur indices and the Schur indices in the presence of certain surface defects.}, while we presumably need to introduce defects in our non-unitary theory in order to compute---directly in 4D---the other Schur indices of the $D_2[SU(2N+1)]$ theories.

As another future direction, we may hope to find information about new observables that are closely related to the chiral algebra as in \cite{Song:2016yfd,Fredrickson:2017yka,Fluder:2017oxm,Imamura:2017wdh}. Moreover, since we have a Lagrangian description of certain Schur observables, it is tempting to see what (if anything) the corresponding correlation functions / OPE coefficients compute in the original strongly interacting theory. Even more simply, it would be interesting to understand if it is possible to map flavor symmetries between our two sets of theories.

Moreover, we expect that our procedure of starting with a unitary 4D theory, mapping to 2D using \cite{Beem:2013sza}, conjugating / permuting the characters, reinterpreting the characters as objects in a unitary 2D theory, and then lifting to a non-unitary theory in 4D will generalize (with certain modifications) to many (and perhaps all) $\mathcal{N}=2$ theories. While we know that the non-unitary 4D theories will not always have a completely free description in terms of hypermultiplets, we expect gauge fields and perhaps the constructions in \cite{Dijkgraaf:2016lym} to play a role (possibly when the original 4D theory has a conformal manifold \cite{Buican:2016arp,Buican:2017uka,Xie:2017vaf,Xie:2017aqx,Choi}). Indeed, we expect non-unitary Lagrangians to be a more diverse and flexible group of objects than their unitary counterparts, and so we expect them to describe \lq\lq more" theories.

Still, we should point out that our non-unitary 4D theories described above have an avatar of 2D unitarity: a modified notion of reflection positivity exists in our theories. Related 2D constructions have played a role in recent work on non-unitary extensions of Zamolodchikov's $c$-theorem \cite{Castro-Alvaredo:2017udm}. Such structures, involving \lq\lq hidden" unitarity, may also shed more light on the question of which non-unitary theories in 2D are able to encode the unitary 4D physics in the original construction of \cite{Beem:2013sza}. We hope to return to this question soon.

It would also be interesting to understand any relation between our construction and the Lagrangians appearing in \cite{Gadde:2015xta,Maruyoshi:2016tqk,Maruyoshi:2016aim,Agarwal:2016pjo,Agarwal:2017roi,Benvenuti:2017bpg,Benvenuti:2017kud,Giacomelli:2017ckh}. While our Lagrangians govern only a particular sector of the theories we study (and perhaps only a particular set of observables in such a sector), they are considerably simpler than the \lq\lq full" Lagrangians in these latter works \footnote{Although, at present, there are no known \lq\lq full" Lagrangians for the $D_2[SU(2N+1)]$ theories with $N>1$.}. Also, our approach is different: we sacrifice unitarity instead of the $\mathcal{N}=2$ superconformal algebra. More generally, it would be interesting to find connections between our discussion and other effective Lagrangian descriptions of sectors of QFTs (e.g., as in \cite{Hellerman:2017sur}).

Finally, we hope to understand if our work is related in any way to supersymmetric localization (see \cite{Pan:2017zie} for some interesting work in this direction from a chiral algebra perspective), to understand if our work is related to the free fermion description of the Schur index for quiver gauge theories \cite{Bourdier:2015sga}, to see how our procedure might work in 6D,  to understand if the theories we have studied here contain some sectors that play a role in the ${\rm dS/ CFT}$ correspondence, and to understand the role our Lagrangians might play in the physics of 3D SCFTs as in \cite{Benvenuti:2017bpg,Benvenuti:2017kud,Aghaei:2017xqe}.

\acknowledgements{We are grateful to T.~Nishinaka for many interesting discussions---including drawing our attention to the formula in \cite{Xie:2016evu} when one of us (M.~B.) visited Kyoto last year---and also for many collaborations on related topics. We also thank S.~Giacomelli and M.~Roberts for correspondence and discussions. M.~B.'s research is partially supported by the Royal Society under the grant \lq\lq New Constraints and Phenomena in Quantum Field Theory." Z.~L. is supported by a Queen Mary University of London PhD studentship.}

\section*{Supplementary Material}
In this appendix, we briefly remind the reader of various formulae and constructions used in the main text:
\begin{itemize}
\item{{\it Schur Operators:} For a more comprehensive review, we refer the reader to \cite{Beem:2013sza,Gadde:2011uv}. Here we simply recall that a Schur operator, $\mathcal{O}$, sits in a short multiplet of the 4D $\mathcal{N}=2$ superconformal algebra (SCA) and is annihilated by the following Poincar\'e supercharges 
\begin{equation}\label{Schurdef}
\left\{\tilde Q_{2\dot-},\CO\right]=\left\{Q^1_{-},\CO\right]=0~,
\end{equation} 
where the numerical indices are spin-half indices of the $su(2)_R\subset u(1)_R\times su(2)_R$ part of the SCA and the sign indices are for the Lorentz group (note that we have dropped $su(2)_R$ and Lorentz indices from $\mathcal{O}$ for simplicity).

\smallskip

The constraints in \eqref{Schurdef} guarantee that the Schur operators have scaling dimensions and $u(1)_R$ charges that are fixed in terms of the $su(2)_R$ and Lorentz weights. Moreover, these operators contribute to the Schur limit of the 4D superconformal index
\begin{equation}\label{SchurIndGen}
I_s(q, x_i)=q^{c_{4d}\over2}\Tr_{\mathcal{H}}(-1)^Fe^{-\beta\Delta}q^{E-R}\prod_ix_i^{f_i}~.
\end{equation}
In this expression, the trace is over the Hilbert space of local operators, $\Delta=\left\{\tilde Q_{2\dot-},\left(\tilde Q_{2\dot-}\right)^{\dagger}\right\}$, $f_i$ are flavor charges, $x_i$ are corresponding flavor fugacities, and the remaining quantities have been described in the main text. In this note, we have simply set $x_i=1$ so that the flavor dependence drops out, and we obtain the unrefined Schur index. Moreover, by the usual arguments of index theory, only states annihilated by $\tilde Q_{2\dot-}$ and $\left(\tilde Q_{2\dot-}\right)^{\dagger}$ contribute to the index (in particular, $\Delta=0$ for these operators).

\smallskip

We will not review the detailed taxonomy of Schur operators here, but we note that any local theory in 4D has a Schur operator: the $su(2)_R$ and Lorentz highest-weight component of the $su(2)_R$ current (this current sits in the same multiplet with the 4D stress tensor). Moreover, any theory in 4D with a local continuous flavor symmetry (in the sense of having a corresponding Noether current) has an additional set of Schur operators: the $su(2)_R$ highest weight components of the corresponding moment maps. This universality of the Schur sector is one of the origins of its power.

Finally, we note that under the chiral algebra mapping alluded to in the main text and explained further in \cite{Beem:2013sza}, we have the following relation between 4D Schur operators and chiral algebra generators
\begin{eqnarray}
\chi\left[J_{+\dot+}^{11}\right]&=&-{1\over2\pi^2}T~, \ \ \chi[\mu^I]={1\over2\sqrt{2}\pi}J^I~, \cr \chi\left[\partial_{+\dot+}\right]&=&\partial_z\equiv\partial~,
\end{eqnarray}
where $J^{11}_{+\dot+}$ is the component of the $su(2)_R$ current described above, $T$ is the 2D holomorphic stress tensor, $\mu^I$ is a moment map of the type we have discussed in the previous paragraph, $J^I$ is an AKM current, $\partial_{+\dot+}$ is a derivative in the $z$ direction of the chiral algebra plane, and $\partial$ is the 2D holomorphic derivative.
}

\item{{\it Modular Quantities:} In the main text we have used several quantities that have nice modular properties. These include the Dedekind $\eta$ function
\begin{equation}
\eta(\tau)=q^{1\over24}\prod_{i=1}^{\infty}(1-q^i)~,
\end{equation}
where $q=e^{2\pi i\tau}$, and the Jacobi theta functions
\begin{eqnarray}
\theta_1(z,\tau)&=&\sum_{n=-\infty}^{\infty}(-1)^{n-{1\over2}}q^{{1\over2}(n+{1\over2})^2}e^{{1\over2}(2n+1)iz}~,\cr\theta_2(z,\tau)&=&\sum_{n=-\infty}^{\infty}q^{{1\over2}(n+{1\over2})^2}e^{{1\over2}(2n+1)iz}~,\cr\theta_3(z,\tau)&=&\sum_{n=-\infty}^{\infty}q^{n^2\over2}e^{2niz}~,\cr\theta_4(z,\tau)&=&\sum_{n=-\infty}^{\infty}(-1)^nq^{n^2\over2}e^{2niz}~.
\end{eqnarray}
In the main text, we have set $z=0$ and defined $\theta_i(0,\tau)\equiv\theta_i(\tau)$.
}

\item{{\it $\widehat{so(2n)}_1$ Characters:} The characters of $\widehat{so(2n)}_1$ are given by \cite{DiFrancesco:1997nk}:
\begin{eqnarray}
\chi_0^{\widehat{so(2n)}_1}&=&{1\over2}\left({\theta_3^n+\theta_4^n\over\eta^n}\right)~,\cr \chi_{1\over2}^{\widehat{so(2n)}_1}&=&{1\over2}\left({\theta_3^n-\theta_4^n\over\eta^n}\right)~,\cr \chi_{{N(N+1)\over4},1}^{\widehat{so(2n)}_1}&=&\chi_{{N(N+1)\over4},2}^{\widehat{so(2n)}_1}=\chi'^{\widehat{so(2n)}_1}_{{N(N+1)\over4}}={1\over2}{\theta_2^n\over\eta^n}~. 
\end{eqnarray}
As mentioned in the main text, when $n=4$ the last three characters are all equal to each other.
}
\end{itemize}

\bigskip


\begin{thebibliography}{}
\bibitem{DiFrancesco:1997nk} 
  P.~Di Francesco, P.~Mathieu and D.~Senechal,
  Springer-Verlag, New York (1997).

\bibitem{Beem:2013sza} 
  C.~Beem, M.~Lemos, P.~Liendo, W.~Peelaers, L.~Rastelli and B.~C.~van Rees,
  Commun.\ Math.\ Phys.\  {\bf 336}, no. 3, 1359 (2015)
  [arXiv:1312.5344 [hep-th]].
  
\bibitem{Gadde:2011uv} 
  A.~Gadde, L.~Rastelli, S.~S.~Razamat and W.~Yan,
  Commun.\ Math.\ Phys.\  {\bf 319}, 147 (2013)
  [arXiv:1110.3740 [hep-th]].
  
\bibitem{Argyres:1995xn} 
  P.~C.~Argyres, M.~R.~Plesser, N.~Seiberg and E.~Witten,
  Nucl.\ Phys.\ B {\bf 461}, 71 (1996)
  [hep-th/9511154].

\bibitem{Cecotti:2010fi} 
  S.~Cecotti, A.~Neitzke and C.~Vafa,
  arXiv:1006.3435 [hep-th].

\bibitem{Buican:2015ina} 
  M.~Buican and T.~Nishinaka,
  J.\ Phys.\ A {\bf 49}, no. 1, 015401 (2016)
  [arXiv:1505.05884 [hep-th]].

\bibitem{Buican:2015hsa} 
  M.~Buican and T.~Nishinaka,
  J.\ Phys.\ A {\bf 49}, no. 4, 045401 (2016)
  [arXiv:1505.06205 [hep-th]].

\bibitem{Cordova:2015nma} 
  C.~Cordova and S.~H.~Shao,
  JHEP {\bf 1601}, 040 (2016)
  [arXiv:1506.00265 [hep-th]].

\bibitem{Buican:2015tda} 
  M.~Buican and T.~Nishinaka,
  JHEP {\bf 1602}, 159 (2016)
  [arXiv:1509.05402 [hep-th]].


\bibitem{RastelliSeminar} 
  L.~Rastelli,
  Harvard University Seminar, November 2014.

\bibitem{Xie:2016evu} 
  D.~Xie, W.~Yan and S.~T.~Yau,
  arXiv:1604.02155 [hep-th].

\bibitem{Kac:2017} 
  V.~G.~Kac and M.~Wakimoto,
  Comptes Rendus Mathematique, {\bf 355}, 128 (2017)
  [arXiv:1612.07423 [math.RT]].

\bibitem{Creutzig:2017qyf} 
  T.~Creutzig,
  arXiv:1701.05926 [hep-th].

\bibitem{Song:2017oew} 
  J.~Song, D.~Xie and W.~Yan,
  arXiv:1706.01607 [hep-th].

\bibitem{Song:2015wta} 
  J.~Song,
  JHEP {\bf 1602}, 045 (2016)
  [arXiv:1509.06730 [hep-th]].

\bibitem{Gaiotto:2009we} 
  D.~Gaiotto,
  JHEP {\bf 1208}, 034 (2012)
  [arXiv:0904.2715 [hep-th]].

\bibitem{Buican:2017fiq} 
  M.~Buican, Z.~Laczko and T.~Nishinaka,
  JHEP {\bf 1709}, 087 (2017)
  [arXiv:1706.03797 [hep-th]].

\bibitem{Buican:2014hfa} 
  M.~Buican, S.~Giacomelli, T.~Nishinaka and C.~Papageorgakis,
  JHEP {\bf 1502}, 185 (2015)
  [arXiv:1411.6026 [hep-th]].

\bibitem{Argyres:2007cn} 
  P.~C.~Argyres and N.~Seiberg,
  JHEP {\bf 0712}, 088 (2007)
  [arXiv:0711.0054 [hep-th]].

\bibitem{Beem:2017ooy} 
  C.~Beem and L.~Rastelli,
  arXiv:1707.07679 [hep-th].

\bibitem{Mathur:1988na} 
  S.~D.~Mathur, S.~Mukhi and A.~Sen,
  Phys.\ Lett.\ B {\bf 213}, 303 (1988).

\bibitem{Mukhi:1989bp} 
  S.~Mukhi and S.~Panda,
  Nucl.\ Phys.\ B {\bf 338}, 263 (1990).

\bibitem{DiPietro:2014bca} 
  L.~Di Pietro and Z.~Komargodski,
  JHEP {\bf 1412}, 031 (2014)
  [arXiv:1407.6061 [hep-th]].

\bibitem{Ardehali:2015bla} 
  A.~Arabi Ardehali,
  JHEP {\bf 1607}, 025 (2016)
  [arXiv:1512.03376 [hep-th]].

\bibitem{Anninos:2011ui} 
  D.~Anninos, T.~Hartman and A.~Strominger,
  Class.\ Quant.\ Grav.\  {\bf 34}, no. 1, 015009 (2017)
  [arXiv:1108.5735 [hep-th]].
  
\bibitem{Hertog:2017ymy}
  T.~Hertog, G.~Tartaglino-Mazzucchelli, T.~Van Riet and G.~Venken,
  arXiv:1709.06024 [hep-th].

\bibitem{Cecotti:2012jx} 
  S.~Cecotti and M.~Del Zotto,
  JHEP {\bf 1301}, 191 (2013)
  [arXiv:1210.2886 [hep-th]].

\bibitem{Cecotti:2013lda} 
  S.~Cecotti, M.~Del Zotto and S.~Giacomelli,
  JHEP {\bf 1304}, 153 (2013)
  [arXiv:1303.3149 [hep-th]].

\bibitem{Wang} 
  C.~Wang and M.~Levin,
  Phys. Rev. Lett.  {\bf 113}, 080403  (2014)
  [arXiv: 1403.7437[cond-mat.str-el]].

\bibitem{Cordova:2017mhb} 
  C.~Cordova, D.~Gaiotto and S.~H.~Shao,
  JHEP {\bf 1705}, 140 (2017)
  [arXiv:1704.01955 [hep-th]].

\bibitem{Pan:2017zie} 
  Y.~Pan and W.~Peelaers,
  arXiv:1710.04306 [hep-th].
  
\bibitem{Song:2016yfd} 
  J.~Song,
  JHEP {\bf 1708}, 044 (2017)
  [arXiv:1612.08956 [hep-th]].

\bibitem{Fredrickson:2017yka} 
  L.~Fredrickson, D.~Pei, W.~Yan and K.~Ye,
  arXiv:1701.08782 [hep-th].
  
\bibitem{Fluder:2017oxm} 
  M.~Fluder and J.~Song,
  arXiv:1710.06029 [hep-th].

\bibitem{Imamura:2017wdh} 
  Y.~Imamura,
  arXiv:1710.08853 [hep-th].


\bibitem{Dijkgraaf:2016lym} 
  R.~Dijkgraaf, B.~Heidenreich, P.~Jefferson and C.~Vafa,
  arXiv:1603.05665 [hep-th].

\bibitem{Buican:2016arp} 
  M.~Buican and T.~Nishinaka,
  J.\ Phys.\ A {\bf 49}, no. 46, 465401 (2016)
  [arXiv:1603.00887 [hep-th]].

\bibitem{Buican:2017uka} 
  M.~Buican and T.~Nishinaka,
  JHEP {\bf 1709}, 066 (2017)
  [arXiv:1705.07173 [hep-th]].

\bibitem{Xie:2017vaf} 
  D.~Xie and S.~T.~Yau,
  arXiv:1701.01123 [hep-th].

\bibitem{Xie:2017aqx} 
  D.~Xie and K.~Ye,
  arXiv:1711.06684 [hep-th].

\bibitem{Choi} 
  J.~Choi and T.~Nishinaka,
  arXiv:1711.07941 [hep-th].

\bibitem{Castro-Alvaredo:2017udm} 
  O.~A.~Castro-Alvaredo, B.~Doyon and F.~Ravanini,
  J.\ Phys.\ A {\bf 50}, no. 42, 424002 (2017)
  [arXiv:1706.01871 [hep-th]].

\bibitem{Gadde:2015xta} 
  A.~Gadde, S.~S.~Razamat and B.~Willett,
  Phys.\ Rev.\ Lett.\  {\bf 115}, no. 17, 171604 (2015)
  doi:10.1103/PhysRevLett.115.171604
  [arXiv:1505.05834 [hep-th]].

\bibitem{Maruyoshi:2016tqk} 
  K.~Maruyoshi and J.~Song,
  Phys.\ Rev.\ Lett.\  {\bf 118}, no. 15, 151602 (2017)
  [arXiv:1606.05632 [hep-th]].

\bibitem{Maruyoshi:2016aim} 
  K.~Maruyoshi and J.~Song,
  JHEP {\bf 1702}, 075 (2017)
  [arXiv:1607.04281 [hep-th]].

\bibitem{Agarwal:2016pjo} 
  P.~Agarwal, K.~Maruyoshi and J.~Song,
  JHEP {\bf 1612}, 103 (2016)
  Addendum: [JHEP {\bf 1704}, 113 (2017)]
  [arXiv:1610.05311 [hep-th]].

\bibitem{Agarwal:2017roi} 
  P.~Agarwal, A.~Sciarappa and J.~Song,
  JHEP {\bf 1710}, 211 (2017)
  [arXiv:1707.04751 [hep-th]].

\bibitem{Benvenuti:2017kud} 
  S.~Benvenuti and S.~Giacomelli,
  arXiv:1706.04949 [hep-th].

\bibitem{Benvenuti:2017bpg} 
  S.~Benvenuti and S.~Giacomelli,
  JHEP {\bf 1710}, 106 (2017)
  [arXiv:1707.05113 [hep-th]].


\bibitem{Giacomelli:2017ckh} 
  S.~Giacomelli,
  arXiv:1710.06469 [hep-th].


\bibitem{Hellerman:2017sur} 
  S.~Hellerman and S.~Maeda,
  arXiv:1710.07336 [hep-th].

\bibitem{Bourdier:2015sga} 
  J.~Bourdier, N.~Drukker and J.~Felix,
  JHEP {\bf 1601}, 167 (2016)
  [arXiv:1510.07041 [hep-th]].

\bibitem{Aghaei:2017xqe} 
  N.~Aghaei, A.~Amariti and Y.~Sekiguchi,
  arXiv:1709.08653 [hep-th].

\end{thebibliography}
\end{document}